\documentclass[aps,prl,showpacs,floatfix,twocolumn,superscriptaddress]{revtex4-2}
\usepackage{graphicx}
\usepackage{bm,amsmath}
\usepackage{dcolumn}
\usepackage{amsfonts,amssymb}
\usepackage{chemfig}
\usepackage{diagbox}
\usepackage{mhchem}
\usepackage{float}

\begin{document}

\title{Ground State of the $S=1/2$ Heisenberg Spin Chain \\ with Random Ferro- and Antiferromagnetic Couplings}

\author{Sibei Li}
\affiliation{School of Physics and Astronomy, Beijing Normal University, Beijing 100875, China}

\author{Hui Shao}
\email{huishao@bnu.edu.cn}
\affiliation{Center for Advanced Quantum Studies, School of Physics and Astronomy, Beijing Normal University, Beijing 100875, China}
\affiliation{Key Laboratory of Multiscale Spin Physics (Ministry of Education), Beijing Normal University, Beijing 100875, China}

\author{Anders W. Sandvik}
\email{sandvik@bu.edu}
\affiliation{Department of Physics, Boston University, 590 Commonwealth Avenue, Boston, Massachusetts 02215, USA}
\affiliation{Beijing National Laboratory for Condensed Matter Physics and Institute of Physics, Chinese Academy of Sciences, Beijing, 100190, China}

 \date{\today}

\begin{abstract}
We study the Heisenberg $S=1/2$ chain with random ferro- and antiferromagnetic couplings using quantum Monte Carlo simulations at
ultra-low temperatures, converging to the ground state. Finite-size scaling of correlation functions and excitation gaps demonstrate
an exotic critical state in qualitative agreement with previous strong-disorder renormalization group calculations but with 
scaling exponents depending on the coupling  distribution. We find dual scaling regimes of the transverse correlations versus the
distance, with an $L$ independent form $C(r)=r^{-\mu}$ for $r \ll L$ and $C(r,L)=L^{-\eta}f(r/L)$ for $r/L > 0$, where $\mu > \eta$ and
the scaling function is delivered by our analysis. These results are at variance with previous spin-wave and density-matrix renormalization
group calculations, thus highlighting the power of unbiased quantum Monte Carlo simulations.
\end{abstract}
 
 \maketitle

Studies of quantum spin chains with random interactions date back to the 1970s, when experimental findings
\cite{Bulaevskii,Theodourou76,Azevedo77,Theodourou77,Bozler80,Tippie81} prompted the development the strong-disorder renormalization
group (SDRG) \cite{Ma79,Dasgupta80}. This method, in turn, led to detailed characterization of the infinite-randomness fixed point
\cite{Fisher94} realized in a wide range of systems \cite{Igloi05, Igloi18};
in particular the $S=1/2$ Heisenberg chain with random antiferromagnetic (AF) couplings.
If ferromagnetic (F) couplings are included (the F-AF chain), a generalized SDRG method generates
large F clusters and the system flows to another fixed point \cite{Furusaki94,Westerberg95,Westerberg97,Hikihara99}.
Because of the approximate nature of the SDRG method in this case, it is not known whether the results represent
the true ground state of the F-AF chain.

Indeed, in recent work a competing quadratic spin wave theory (SWT) was put forward \cite{Fava23} according to which the ground state is ordered
for any spin $S$ (which had not been addressed in SWT calculations at temperature $T>0$ \cite{Wan02}). This scenario was supported by
density matrix renormalization group (DMRG) calculations for $S=1/2$, $1$, and $3/2$, which indicated that the ground state is
ordered (with a ``staggered'' order parameter), though with large finite-size effects for $S=1/2$.

Motivated by the competing theories for this important paradigmatic random-coupling system, which also has an experimental realization
\cite{Nguyen96}, we here revisit the $S=1/2$ F-AF chain. The DMRG method is often impeded by convergence
problems for random systems \cite{Rapsch99,Goldsborough15,Lin17,Tsai20,Wada22,Lin23,Fava23}, and we here employ quantum Monte Carlo (QMC)
calculations with the Stochastic Series Expansion (SSE) method at ultra-low temperatures, reaching the ground state for systems almost
twice larger than in Ref.~\onlinecite{Fava23} ($L=144$ versus $L=80$) and with more favorable periodic boundary conditions. With many results
deviating from SWT predictions, we conclude that an ordered state is unlikely. We find generally good agreement with SDRG predictions
\cite{Furusaki94,Westerberg95,Westerberg97,Hikihara99}, though some exponents depend on the coupling distribution. We also characterize
the exotic ground state beyond previous SDRG results.

\begin{figure}[t]
\includegraphics[width=75mm]{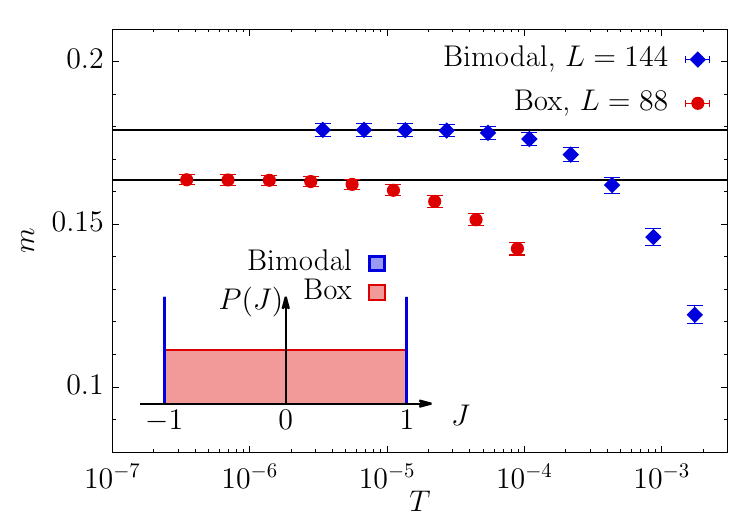}
\caption{Convergence of the disorder averaged order parameter Eq.~(\ref{mst}) with lowered $T=L^{-1}2^{-n}$ (with positive integer $n$)
for the largest systems with couplings drawn from the bimodal (blue diamonds) and box  (red circles) distributions. The distributions are
illustrated in the inset and the maximum coupling $|J|=1$ is the unit of $T$.}
\label{dis}
\end{figure}

{\it Model and QMC method}---In the $S=1/2$ F-AF Hamiltonian 
\begin{equation}
 H=\sum_{i=1}^LJ_i\mathbf{S}_i \cdot \mathbf{S}_{i+1},~~~\mathbf{S}_{L+1}=\mathbf{S}_{1},
\label{ham}
\end{equation}
we draw $J_i$ from one of the distributions in the inset of Fig.~\ref{dis}. The bimodal distribution was studied in Ref.~\onlinecite{Fava23}
and we also consider the ``box'' distribution. With periodic boundaries, we impose an even number of $J_i > 0$ instances in order to
avoid the QMC sign problem \cite{Henelius00}.

Previous QMC simulations \cite{Frischmuth97,Frischmuth99} focused on thermodynamics at low but nonzero $T$ and in general found good agreement
with SDRG predictions. With the SSE method, the ground state can be reached for sufficiently low $T$, as in many previous studies of random
quantum spin models \cite{Sandvik02,Laflorencie04,Laflorencie06,Wang06,Shu16,Dao24}. For the F-AF chain we find short equilibration times,
thus allowing for a large number of disorder samples; here typically 1000-2000. As shown in Fig.~\ref{dis}, we reach the ground state for
$L \le 144$ with the bimodal distribution. Because of typically smaller excitation gaps with the box distribution (which includes $J=0$)
the convergence is slower and $L \le 88$ in this case. The SSE computational effort scales as $L/T$, and we use tests such as Fig.~\ref{dis}
to judge the convergence of all quantities studied.

The ground state typically has non-zero total spin. For a given sample $\{J_i\}$, it is useful to
introduce an accumulated phase for each {\it site} $i$ based on the signs,
\begin{equation}
c_i=\prod_{k=1}^i{\rm sgn}(-J_k),
\end{equation}
so that $H$ is classically minimized for $S^z_i = c_i/2$. With $n_{\rm A}$ and $n_{\rm B}$ denoting the number of $c_i=+1$ and $c_i=-1$ sites, respectively,
the ground state has total spin $S_{\rm tot}=|n_{\rm A}-n_{\rm B}|/2$ \cite{Lieb62}; thus $S_{\rm tot}$ is typically of order $L^{1/2}$. It is this growth
of $S_{\rm tot}$ with $L$ that potentially could enable long-range order, similar to a ferromagnet where $S_{\rm tot}=L/2$, without violating the
order-forbidding Mermin-Wagner theorem \cite{Mermin66}. We here exclude samples with $n_{\rm A}=n_{\rm B}$, so that $S_{\rm tot}>0$ always holds.

{\it Order parameter}---Working in the $z$ basis, it is convenient to define the order parameter in the sector with
$S^z=\sum_i S^z_i=S_{\rm tot}$ as \cite{Frischmuth99,Fava23}
\begin{equation}
m(L)=\frac{1}{L} \sum_{i=1}^L [\langle c_iS_i^z \rangle ],~~~S^z=S_{\rm tot}=|n_{\rm A}-n_{\rm B}|/2,
\label{mst}
\end{equation}
where $\langle\rangle$ is the expectation value for an individual sample and $[]$ denotes sample averaging. In our SSE simulations $S^z$ fluctuates among
$\{-S_{\rm tot},S_{\rm tot}\}$ with equal probability, and we group all measurements accordingly. We also analyze the longitudinal correlation function
\begin{equation}
C_{\parallel}(r) = [\langle c_ic_{i+r}S_i^zS_{i+r}^z \rangle ],~~~S^z=S_{\rm tot}.
\label{clong}
\end{equation}
Whether or not $m(L)$ remains non-zero when $L \to \infty$ is the main point of contention \cite{Furusaki94,Hikihara99,Fava23}.
If there is order, then $C_{\parallel}(r\to \infty) = m^2(L \to \infty)$.

\begin{figure}[t]
\includegraphics[width=75mm]{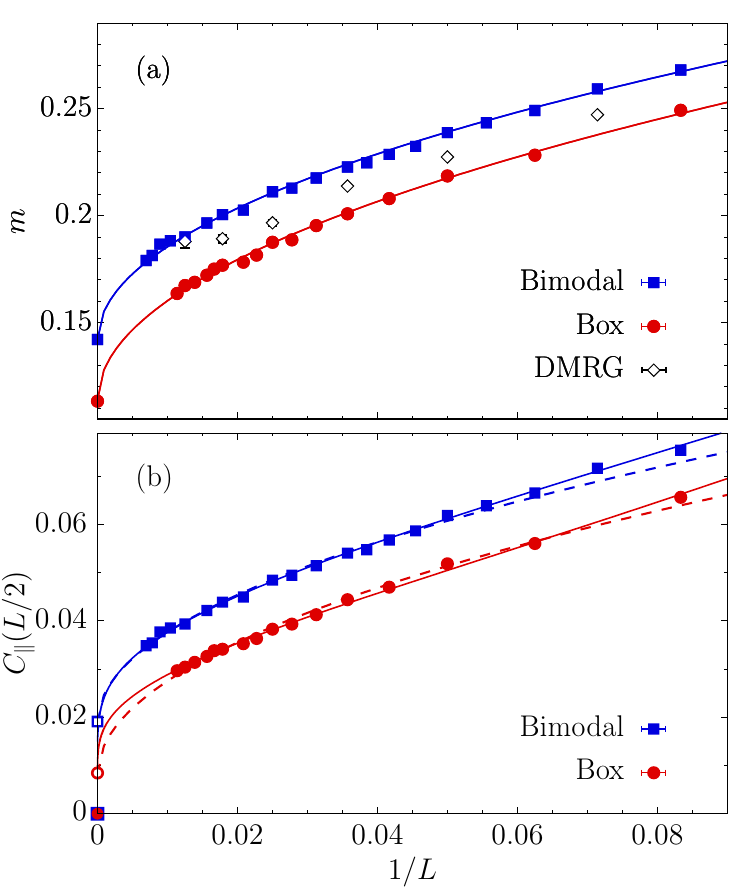}
\caption{(a) The order parameter Eq.~(\ref{mst}) and (b) the long-distance longitudinal correlation function Eq.~(\ref{clong}), both vs the
inverse system size. Results for the bimodal and box distributions are shown with blue squares and red circles, respectively. The solid
curves show fits $m=b + aL^{-1/2}$ in (a) and $C_{\parallel}=a\ln^{-1}(bL)$ in (b), expected from SWT and SDRG, respectively, with $a$ and $b$
optimized. In (b) the dashed curves show $C_{\parallel}=b + aL^{-1/2}$ fits. The diamonds in (a) show the DMRG results for
open chains \cite{Fava23}.}
\label{order}
\end{figure}

\begin{figure}[H]
	\includegraphics[width=75mm]{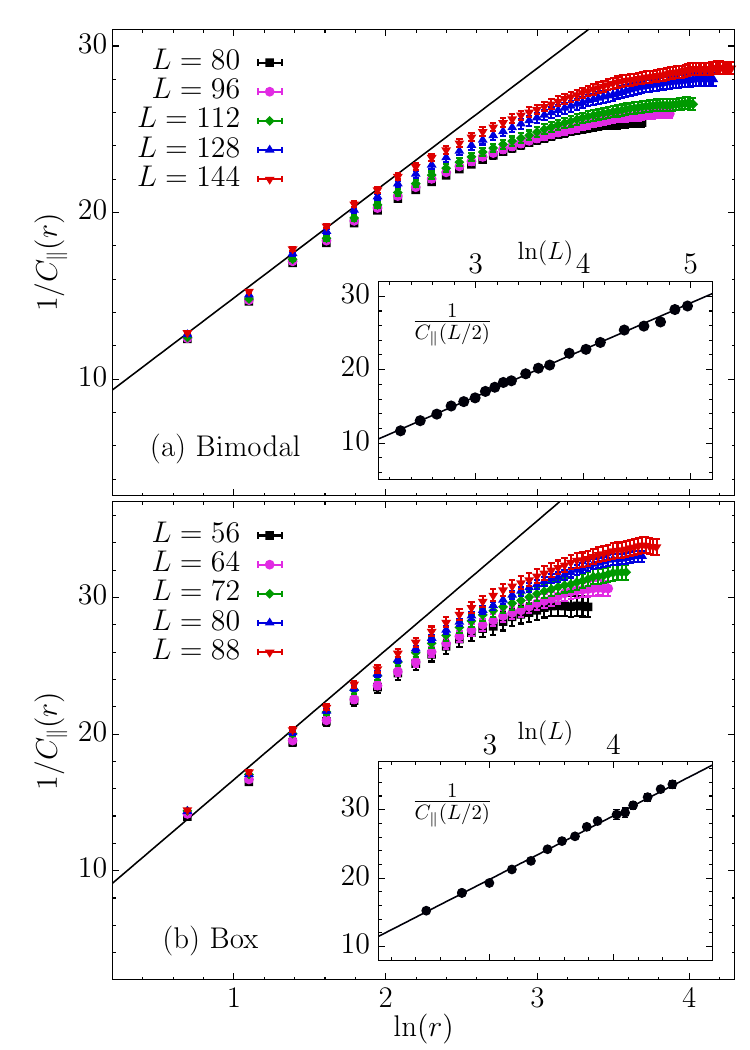}
	\vskip-3mm
	\caption{The inverse longitudinal correlation function Eq.~(\ref{clong}) graphed vs $\ln(r)$ for different system sizes, for the bimodal distribution
		in (a) and box distribution in (b). The inset shows the $r=L/2$ data vs $\ln(L)$, where the slopes of the lines corresponding to the form
		$C_\parallel \sim \ln^{-1}(r/r_0)$ are determined and fixed to the same values in the main plots.}
	\label{lc2}
\end{figure}

In Figs.~\ref{order}(a) and \ref{order}(b) we graph the order parameter and the $r=L/2$ correlation function, respectively, versus $1/L$.
The order parameter can be fitted to the SWT form, $m(L) \sim m(\infty)+aL^{-1/2}$ \cite{Fava23}, but the correlation function
(which typically is a more sensitive detector of long-range order) in Fig.~\ref{order}(b) matches the behavior in numerical
SDRG, $C_\parallel(r) \propto \ln^{-1}(r/r_0)$ \cite{Hikihara99}; see also Fig.~\ref{lc2} in the End Matter. The correlation function can
also be fitted under the assumption of long-range order with $m^2(\infty)$ in reasonable agreement with Fig.~\ref{order}(a), though the SDRG
form works marginally better for the largest systems in the case of box distribution. Overall, neither theory can be ruled out by the
results in Fig.~\ref{order} alone.

In Fig.~\ref{order}(a) we also show the DMRG results \cite{Fava23}. Here the data flatten out for the largest systems, seemingly providing a
stronger case for an ordered state but conflicting with the SWT predicted $L^{-1/2}$ correction. The most likely explanation for the observed
behavior is that the DMRG results for the largest systems are not fully converged, despite best efforts to control errors. The behavior for the
smaller systems is similar to our results, with the overall shift likely a consequence of the different boundary conditions. 

{\it Transverse correlations}---The transverse correlation function can be defined in analogy to Eq.~(\ref{clong}) using the $x$ components of the
spins \cite{Fava23}, which can also be computed with SSE but at higher computational cost than the $z$ component \cite{Dorneich01}. Here we take
a different approach that is equivalent for $L \to \infty$ since $S_{\rm tot}$ grows with $L$; using the $z$ component in the sector with $S^z=0$:
\begin{equation}
C_{\perp}(r) = [ \langle c_ic_{i+r}S_i^zS_{i+r}^z \rangle ],~~~S^z= 0.
\label{ctrans}
\end{equation}
In the $S^z=0$ sector the order parameter falls in the $xy$-plane, thus $C_{\perp}(r) \to 0$ for $r \to \infty$ even if the system is ordered
and $C_{\perp}(r) \not= C_{\parallel}(r)$ even if there is no long-range order. The SWT prediction here is a universal $r^{-1/2}$ decay, but we
are not aware of any SDRG prediction. 

We only consider the average correlations. The often studied definition ${\rm exp}[\ln(C_{ij})]$ of the typical correlation
function is slightly biased when the nonlinear operation on the mean correlator $C_{ij}$ of spins $i$, $j$ is taken with noisy data before
spatial and disorder averaging.

\begin{figure}[t]
\includegraphics[width=75mm]{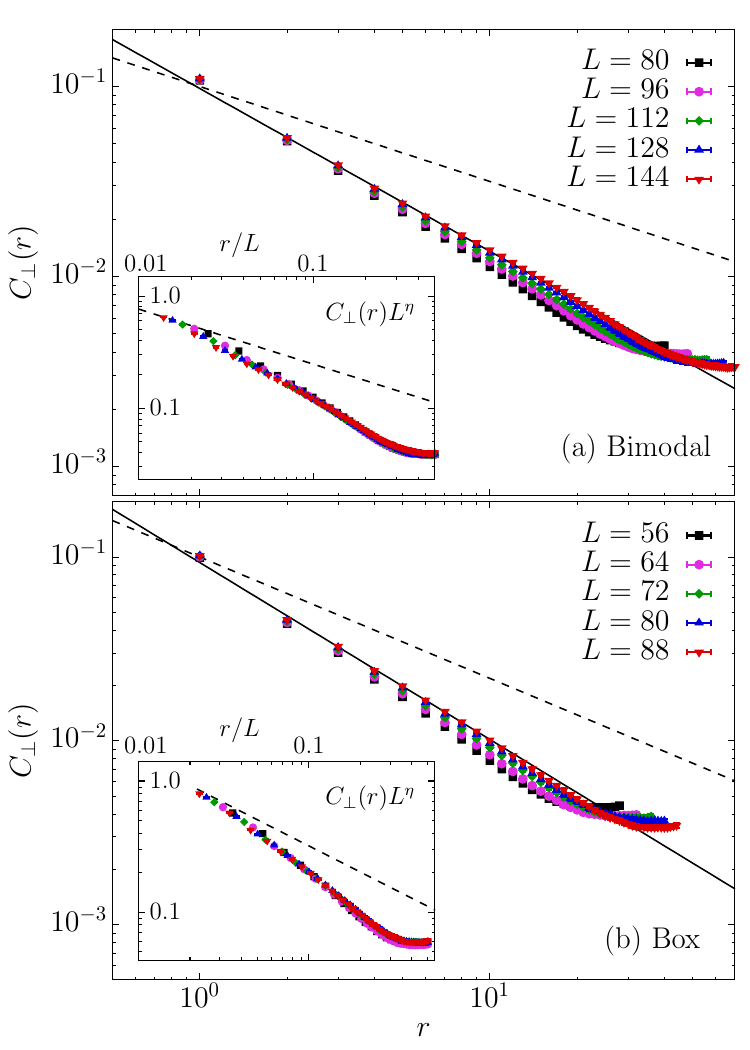}
\caption{Transverse correlation function Eq.~(\ref{ctrans}) for (a) the bimodal and (b) the box distribution, in each case
for five system sizes. The solid lines show the behavior $r^{-\mu}$, with $\mu = 0.85$ in (a) $\mu=0.95$ in (b). The insets
show the same data vs $r/L$ for $r \ge 2$ multiplied by $L^\eta$, with $\eta=0.49$ in (a) and $\eta=0.66$ in (b).
In all plots, the dashed line shows the power law with $\mu=\eta$, which in (a) coincides closely with the SWT prediction $\mu=1/2$.}
\label{corr}
\end{figure}

\begin{figure}[t]
	\includegraphics[width=75mm]{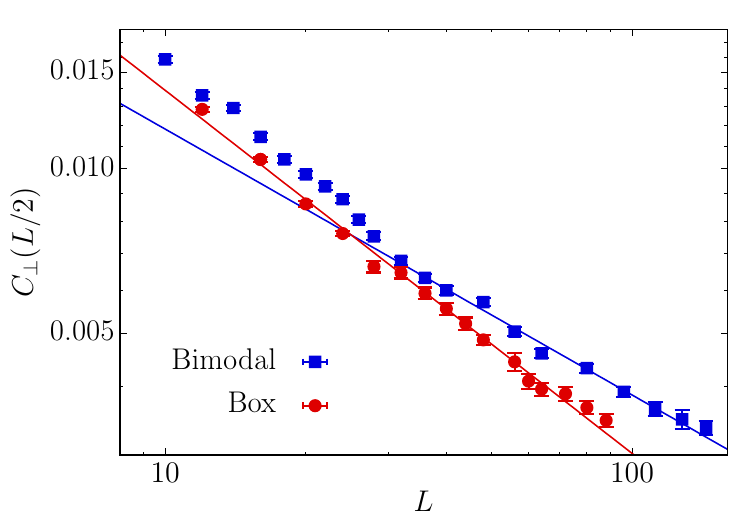}
	\vskip-3mm
	\caption{Transverse correlation function at $r=L/2$ vs the system size $L$ for the bimodal (blue squares) and box (red circles)
		coupling distributions. The lines show fits of the form $C_\perp \propto L^{-\eta}$, with $\eta =0.49 \pm 0.02$ (bimodal distribution) and
		$\eta =0.66 \pm 0.02$ (box distribution).}
	\label{corl2}
\end{figure}

As shown in Fig.~\ref{corl2} in the End Matter,
our results for $C_{\perp}(r=L/2)$ exhibit a decay $L^{-\eta}$ with $\eta=0.49 \pm 0.02$, seemingly in
agreement with the SWT prediction $\eta=1/2$ (though a logarithmic correction was also predicted for $r \to L/2$ \cite{Fava23}).
For the box distribution the decay is faster, with $\eta = 0.66 \pm 0.02$. In both cases, however, the full behavior versus $r$ is much more intricate,
 as shown in Fig.~\ref{corr}. The $r\ll L$ data suggest convergence to a power law $r^{-\mu}$, with $\mu = 0.85 \pm 0.01$ and
$\mu = 0.95 \pm 0.01$ for the bimodal and box distributions, respectively, obtained by fits to a range of $r>3$ for the largest system
sizes in each case. Though the exponents may change slightly for larger $L$, it is hard to imagine that the $L \to \infty$ behavior
would agree with the SWT form $C_{\perp}\propto r^{-1/2}$ (for the bimodal distribution).

Since $\mu > \eta$ and the overall $L^{-\eta}$ decay applies when $r \to L/2$, the putative $r \ll L$ power law $r^{-\mu}$ can only hold up to
$r \propto L^{\mu - \eta}$; otherwise a minimum in $C_\perp (r)$ has to develop, of which we see no sign. In the insets of Fig.~\ref{corr} we
show the correlation functions scaled by $L^\eta$ and graphed versus $r/L$. Here we observe data collapse for the larger systems, suggesting
that the correlation function is of the form $C_{\perp}(r) \propto L^{-\eta}f(r/L)$ for all $r/L>0$, which is still compatible with an
$L$ independent power law $r^{-\mu}$ for $r < L^{\mu -\eta}$. In the case of the longitudinal correlations, as discussed in Fig.~\ref{lc2} in the End Matter, 
the $r \ll L$ and $r=L/2$ behaviors follow the same $\ln^{-1}(r)$ form.

The transverse correlations were not studied with SDRG, and its unusual scaling is another
sign of an exotic ground state induced by the random-sign couplings, which force ground states with $S_{\rm tot}>0$. If the SWT applies instead,
we should have $\mu=\eta=1/2$ (for the bimodal case, and $\mu=\eta$ also more generally for pure $r$ dependence), and the collapsed function
$f(r/L)$ in Fig.~\ref{corr}(a) would have to change considerably, despite almost absent finite-size corrections with the present system sizes.

The DMRG computed $C_\perp(r=L/2)$ \cite{Fava23} also appears to be described by $\eta=1/2$, and the behavior for small $r$ also looks overall similar
to our results here, despite the different boundary conditions and slightly different definitions of $C_\perp$. It was suggested that the behavior
for a very large system will eventually agree with the SWT, but there is neither numerical evidence for slow convergence nor predictions of specific
large scaling corrections.

\begin{figure}[t]
\includegraphics[width=75mm]{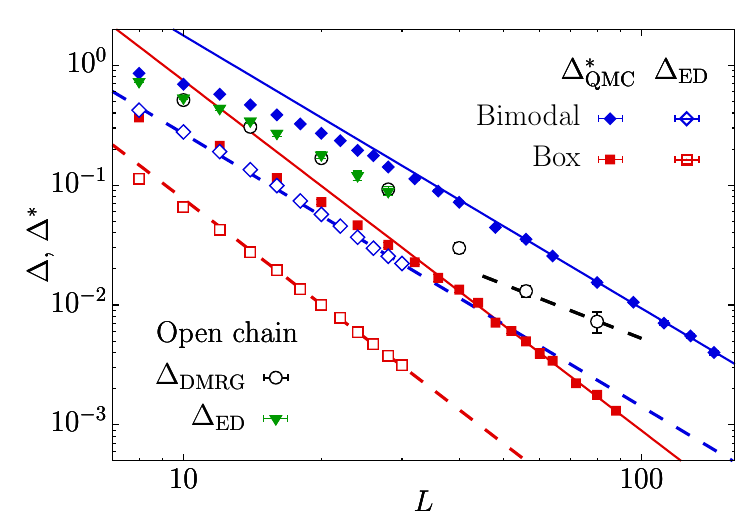}
\caption{Lanczos ED gaps for periodic $L \ge 30$ chains (open red and blue symbols) compared with the upper bound
Eq.~(\ref{deltastar}) (solid red and blue symbols) for the bimodal (blue) and box (red) distributions. Fits to the $\Delta^*_{\rm QMC}$
data give the dynamic exponent $z=2.27 \pm 0.03$ and $z=3.00 \pm 0.09$ for the bimodal and box distribution, respectively. Lines with the
same slopes are drawn through the ED data, showing compatibility with the same exponents. The DMRG results for open chains \cite{Fava23}
(bimodal distribution) are shown with the black open circles and are compared to our ED results (green triangles), using the same gap definition.
The short long-dashed black line shows the SWT prediction $z=3/2$.}
\label{gaps}
\end{figure}

\begin{figure}[t]
	\includegraphics[width=75mm]{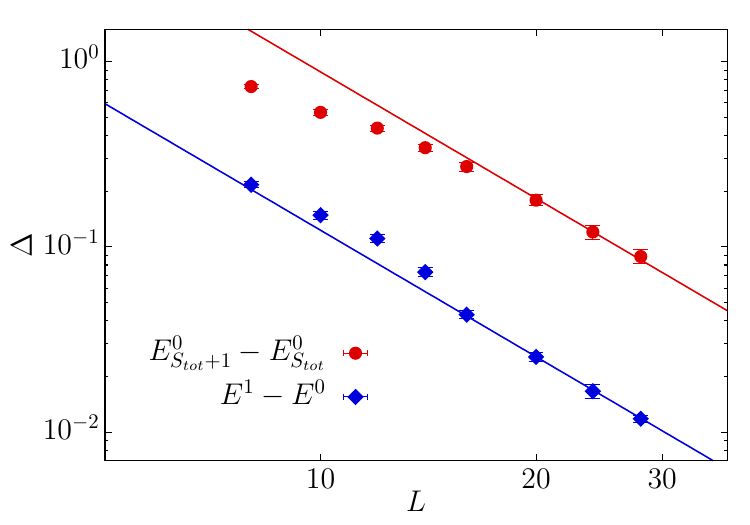}
	\vskip-3mm
	\caption{Comparison of the gap definition based on the lowest state with $S_{\rm tot}+1$ \cite{Fava23} and the true smallest gap
		$E^1-E^0$, where $E^1$ is the lower of the ground state energies in the $S_{\rm tot}-1$ or $S_{\rm tot}+1$ sector. These results are for open
		chains with bimodal coupling distribution, and the lines correspond to the dynamic exponent $z=2.27$ extracted in Fig.~4 in the main Letter.}
	\label{gaptest}
\end{figure}

{\it Excitation gap}---Next we study the disorder averaged gap $\Delta(L)$, which should depend on the dynamic exponent $z$; $\Delta(L) \sim L^{-z}$.
In the DMRG calculations \cite{Fava23} the gap for a system with spin $S_{\rm tot}$ was computed in the $S_{\rm tot}+1$ sector. However, the smallest
gap is roughly half of the times in the $S_{\rm tot}-1$ sector, and we here extract the smallest of the two gaps using Lanczos exact diagonalization.
In Fig.~\ref{gaptest} in the End Matter, we show that the $S_{\rm tot}+1$ and $S_{\rm tot}-1$ gaps scale in the same way, as expected.

\begin{figure}[H]
	\includegraphics[width=75mm]{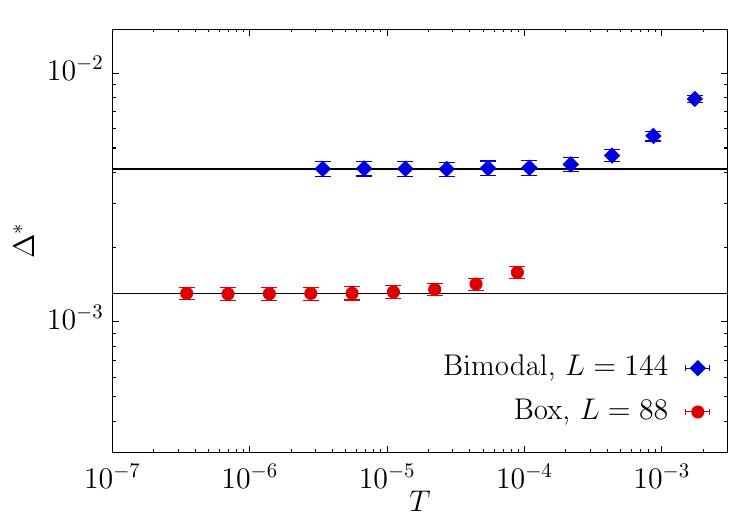}
	\vskip-3mm
	\caption{Disorder averaged gap upper bounds according to Eq.~(\ref{deltastar}), graphed versus the temperature to illustrate
		the convergence of this quantity in the same way as the order parameter in Fig.~\ref{dis}.}
	\label{converge}
\end{figure}

In addition, to exact disorder averaged gaps for $L \le 30$, we also use SSE for an upper bound obtained from well known sum rules \cite{Wang06};
\begin{equation}
\Delta^*=2\left [ \frac{S_{\rm st}}{\chi_{\rm st}} \right ] \geqslant \Delta,
\label{deltastar}
\end{equation}
where $S_{\rm st}$ is the generalized staggered structure factor,
\begin{equation}
S_{\rm st}=\frac{1}{L} \left \langle \left (\sum_{i=1}^L c_iS_i^z \right )^2 \right \rangle,
\label{Sst}
\end{equation}
and ${\chi_{\rm st}}$ is the corresponding susceptibility, 
\begin{equation}
\chi_{\rm st}=\frac{1}{L} \sum_{i,j=1}^{L}c_ic_j\int_{0}^{\beta}\langle S_i^z(\tau)S_j^z(0)d\tau\rangle.
\label{chist}
\end{equation}
Note that, the ratio in Eq.~(\ref{deltastar}) is taken before disorder averaging. We demonstrate convergence
of $\Delta^*$ with lowered $T$ in Fig.~\ref{converge} in the End Matter.

Results for both exact and upper-bound gaps are shown in Fig.~\ref{gaps}. Here we indeed observe power law decays, with only small
finite-size corrections in the ED data. The upper bounds are much larger than the true gaps, but the scaling is the same for the larger systems.
This asymptotic agreement is expected, as the entire low-energy spectrum should exhibit scaling with $z$. We find different dynamic exponents
for the two distributions. In the case of the bimodal distribution, the estimate $z=2.27 \pm 0.03$ is close to the value obtained with SDRG
\cite{Westerberg95,Westerberg97,Igloi05},
in that case for several different coupling distributions. There is no visible trend in our data of bringing the exponent to the SWT
predicted value $3/2$ eventually. For the box distribution the exponent is larger still, $z=3.00 \pm 0.09$, in apparent conflict 
with the universal SDRG value.

We also plot the previous DMRG gaps in Fig.~\ref{gaps} and compare with our own ED calculations for $L \le 30$ open chains.
Here we have also used the definition of the gap to the lowest $S_{\rm tot}+1$ level. The agreement between the DMRG and ED results is
good for the systems where both are available. The DMRG results are also largely compatible with our $z$ estimate if we assume that
the results for the largest systems are affected by minor convergence problems. If we instead take the data at face value as numerically
exact, it would appear that the large-$L$ behavior agrees with the SWT exponent. However, we find such a rapid change in decay form
unlikely.

{\it Discussion}---The non-zero typical spin $S_{\rm tot} \propto \sqrt{L}$ implies different longitudinal and transverse correlation functions, but
this scaling of the spin is not necessarily sufficient for inducing long-range order, unlike the trivial ferromagnet where $S_{\rm tot}=L/2$. However,
if there is long-range order, at least some fraction of it must be induced in the spin-$z$ components when $S^z=S_{\rm tot}$, as seen in an extreme
case of an $S_{\rm tot}=1/2$ ground state of a two-dimensional system with an impurity \cite{Sanyal12}.

Given that the longitudinal correlations in Figs.~\ref{order}(b) and \ref{lc2} show consistency with the same inverse logarithmic $L$ dependence as in 
numerical SDRG \cite{Hikihara99} and that all other results presented here disagree with the SWT, the SDRG scenario is most likely correct.
In the case of the bimodal distribution we even observe quantitative agreement with the SDRG dynamic exponent. As a matter of principle, our results
for the order parameter and correlations in Fig.~\ref{order} cannot alone exclude long-range order by some unknown alternative mechanism. However, ordered
quantum magnets are universally believed to be described by SWT, and the invalid predictions for the F-AF chain then speak against an ordered state.

The differences between the bimodal and box distribution suggest nonuniversal finite-randomness fixed points for the F-AF chain. We have furthermore
demonstrated a unique dual scaling behavior of the transverse correlations that was not predicted by the SDRG. While a power law $C_\perp \sim r^{-\mu}$
develops for all $r \ll L$, the size dependence remains for $r/L > 0$ even when $L \to \infty$, in the form $C_\perp \sim L^{-\eta}f(r/L)$. This behavior as
well should be a direct consequence of $S_{\rm tot} \propto \sqrt{L}$. 
With no order but $C_\perp$ and $C_\parallel$ decaying in different ways, the ground state can be regarded as a nematic.

Looking back at the DMRG calculations \cite{Fava23}, $C_\perp(r)$ even for $S=1$ is not close to the SWT prediction, while for $S=3/2$ the
short-distance behavior is quite close but only for $r \in [2,4]$. Even in the absence of long-range order, for large $S$ the SWT predictions
should still hold up to some distance, and the behavior for $S=3/2$ could possibly reflect crossover from the SWT form to a decay to zero
in the disordered ground state that we have demonstrated here for $S=1/2$. This behavior may possibly persist for all $S < \infty$, though
a critical $S$ above which the system orders cannot be ruled out.

\begin{figure}[t]
	\includegraphics[width=75mm]{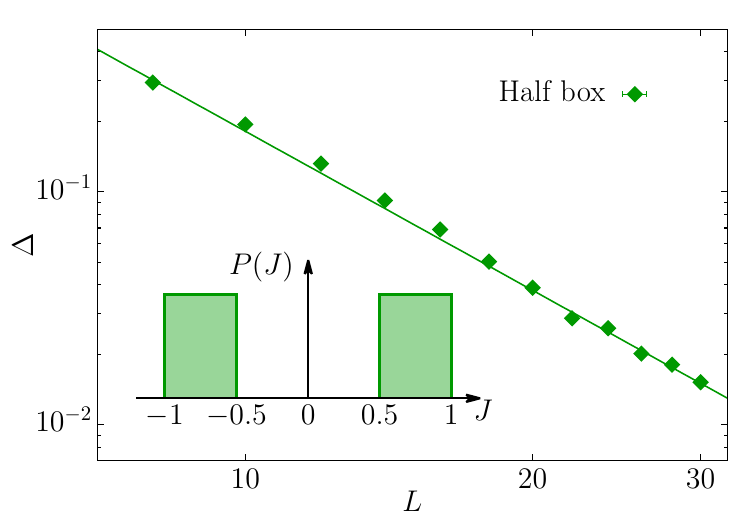}
	\vskip-3mm
	\caption{Gap scaling for periodic chains with a ``half box'' coupling distribution, illustrated in the inset. The line corresponds to power law
		behavior with the same dynamic exponent as for the bimodal distribution in Fig.~4 in the main paper; $z=2.27$.}  
	\label{halfbox}
\end{figure}

\begin{figure}[t]
	\includegraphics[width=75mm]{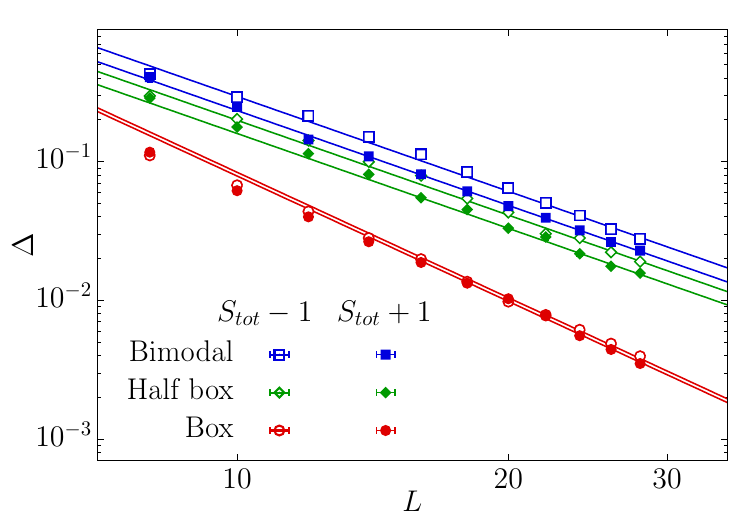}
	\vskip-3mm
	\caption{Disorder averaged ED gaps between the ground state with spin $S_{\rm tot}$ and those in the sectors $S_{\rm tot}+1$ and $S_{\rm tot}-1$
		for the bimodal (blue squares), half box (green diamonds) and full box (red circles) distributions. The lines correspond to power law behavior
		with $z=2.27$ for the bimodal and half-box distribution and $z=3.00$ for the full box distribution.}
	\label{differents}
\end{figure}

In Figs.~\ref{halfbox} and \ref{differents} in the End Matter, 
we also show ED results for the gap scaling in the case of a ``half box'' distribution $P(J)$,
with uniform $|J| \in [0.5,1]$. The scaling is fully consistent with the same dynamic exponent as with the bimodal distribution, with
no apparent differences in the rate of convergence to the asymptotic power law. Based on the results for all three coupling distributions, we therefore
conjecture that there are two universality classes of the F-AF chain, for distributions with and without weight at $J=0$ (and possibly other
classes for singular distributions). Critical phases with varying exponents have previously been found in two-dimensional quantum spin models,
e.g., in Refs.~\onlinecite{Liu18,Liu20}, and also there with a competing proposal of an ultimately ordered ground state \cite{Kimchi18}. Further work
with different coupling distributions should be carried out to further elucidate the degree of non-universality in the F-AF chain.

\begin{acknowledgments}

{\it Acknowledgments}---We would like to thank Akshat Pandey
 for discussions prompting us to undertake this project and for many valuable comments.
 We also thank Adam Nahum for useful discussions.
This work was supported by 
the National Natural Science Foundation of China under Grant No.~12122502,  
National Key Projects for Research and Development of China under Grant No.~2021YFA1400400, 
the Fundamental Research Funds for the Central Universities (to H.S.), 
and by the Simons Foundation under Grant No.~511064 (to A.W.S.). 
Some of the numerical calculations were carried out on the Shared Computing Cluster managed by 
Boston University's Research Computing Services.
\end{acknowledgments}

\onecolumngrid
\vspace{0.3cm}
\begin{center}
\large\bf{End Matter}
\end{center}
\vspace{0.3cm}
\twocolumngrid
	
{\it Longitudinal correlations}---In Fig.~\ref{lc2} we graph the inverse longitudinal correlation function versus $\ln(r)$ for different system sizes, as done for SDRG data
in Ref.~\onlinecite{Hikihara99}. Though we do not have sufficiently large system sizes to converge to a clear $\ln^{-1}(r/r_0)$ behavior
for large $r$, the results are certainly consistent with such behavior with increasing $L$. For $r=L/2$ the behavior
$\ln^{-1}(L/L_0)$, shown in the insets of Fig.~\ref{lc2}, is essentially perfect [as already seen in Fig.~\ref{order}(b)]. 

{\it Long-distance transverse correlations.}---Figure \ref{corl2} shows results for the transverse correlation function at distance $r=L/2$. After a cross-over from non-asymptotic
behavior, the results for both coupling distributions exhibit good power-law scaling, $C_\perp \propto L^{-\eta}$, with no sign of further
significant corrections. The exponent $\eta$ depends on the distribution. A fit to the data for $L \in [40,144]$ gives $\eta =0.49 \pm 0.02$
for the bimodal distribution, while for the box distribution $\eta =0.66 \pm 0.02$ when fitting to data for $L \in [15,88]$.

Interestingly, for the smaller sizes both data sets show the same rate of decay. This behavior can naively be explained by
the fact that small systems under the box distribution will only rarely contain couplings very close to $J=0$, 
and thus the system with bimodal distribution
is typically not that different. However, as the system size increases, most samples drawn from the box distribution will have some small
couplings; thus the systems can behave differently. This argument does not tell us which one of the system should show a cross-over behavior, which
is clearly the bimodal one in Fig.~\ref{corl2}. The box distribution gives almost the asymptotic decay already from very small system sizes, while the
bimodal system shows a rather sharp crossover around $L=30$.

In Ref.~\onlinecite{Fava23}, the SWT predicts $C_{\perp}(L/2) \sim \ln(L)L^{-1/2}$ for the bimodal distribution, where the multiplicative logarithm
would cause an effectively smaller exponent $\eta$ (slower decay) for small $L$, opposite to what we observe in Fig.~\ref{corl2}. Thus, we conclude
that our results do not support the SWT scenario in this regard. Note that the log correction arises from rare events and was only predicted when
$r \to L$, with the pure power law $C_{\perp}(r) \sim r^{-1/2}$ pertaining away from this limit.

{\it Gap comparisons.}---Fava et al.~analyzed only the gap between the ground state with spin $S_{\rm tot}$ and the lowest state in the sector $S_{\rm tot}+1$ \cite{Fava23}.
In our ED study we found that the lowest gap has $S_{\rm tot}+1$ or $S_{\rm tot}-1$ with the fraction of each approaching $0.5$ with increasing $L$.
Though we fully expect the gaps defined in either sector to scale with the same $z$ asymptotically, it is still interesting to investigate
both definitions. To compare directly with Ref.~\onlinecite{Fava23}, we first consider open chains and show the results in Fig.~\ref{gaptest}. Here
we indeed find the same scaling for the larger systems, but the mean gap when only using the $S_{\rm tot}+1$ sector is almost an order of
magnitude larger than our definition based on the smallest gap.

To complement our studies of the bimodal and box coupling distributions, we have also carried out ED calculations of the gap for
a ``half box'' distribution, illustrated in the inset of Fig.~\ref{halfbox}. Here in the main part of the figure we find that the average
gap scales in a way fully consistent with that for the bimodal distribution. Considering also our results for the full box distribution,
the good agreement suggests that distributions $P(J)$ with weight at $J=0$ behave differently. In principle, if the fixed point from the SDRG
calculations applies, one may expect slow convergence to a singular distribution when the microscopic distribution has no weight at $J=0$.
However, since the SDRG procedure for the F-AF chain does not flow to the IRFP, the final distribution should not be extremely singular. Given also the
fact that there are no apparent difference in the convergence rate between the full and half box distributions, as shown in Fig.~\ref{differents}
for periodic chains, we find it more likely that distributions with and without weight at $J=0$ fall into (at least) two distinct classes of fixed points.

{\it Convergence of the gap bound}---In Fig.~\ref{dis} we demonstrated only the convergence of the order parameter. As another important example, we show in
Fig.~\ref{converge} that the upper bound of the gap Eq.~(\ref{deltastar}) converges even faster.

\end{document}